\documentclass{ieeeaccess}

\usepackage{graphicx}
\usepackage{latexsym}
\usepackage{amsmath}
\usepackage{amsthm}

\usepackage{amsfonts}
\usepackage{subfigure}
\usepackage{dsfont}
\usepackage{epstopdf}
\usepackage{threeparttable}
\usepackage{multirow}

\usepackage{enumerate}

\usepackage{epsfig}

\usepackage{bm}

\usepackage{cite}

\usepackage{algorithmic}

\usepackage{textcomp}

\def\BibTeX{{\mathrm B\kern-.05em{\sc i\kern-.025em b}\kern-.08em
    T\kern-.1667em\lower.7ex\hbox{E}\kern-.125emX}}

\begin{document}
\history{Date of publication 0000 00, 0000, date of current version 0000 00, 0000.}
\doi{XXXXXX }

\title{ELM-based Frame Synchronization in Nonlinear Distortion Scenario Using Superimposed Training}
\author{\uppercase{Chaojin Qing}\authorrefmark{1}, \IEEEmembership{Member, IEEE},
\uppercase{Wang Yu}\authorrefmark{1}, \uppercase{Shuhai Tang}\authorrefmark{1},
\uppercase{Chuangui Rao}\authorrefmark{1},
\uppercase{and Jiafan Wang}\authorrefmark{2}}
\address[1]{School of Electrical Engineering and Electronic Information, Xihua University, Chengdu, 610039, China. (e-mail: qingchj@uestc.edu.cn)}
\address[2]{Synopsys Inc., 2025 NE Cornelius Pass Rd, Hillsboro, OR 97124, USA.}

\tfootnote{This work is supported in part by the Key Projects of Education Department of Sichuan Province (Grant 15ZA0134), the Special Funds of Industry Development of Sichuan Province (Grant zyf-2018-056), the Major Special Funds of Science and Technology of Sichuan Science and Technology Plan Project (Grant No. 19ZDZX0016 /2019YFG0395), the Demonstration Project of Chengdu Major Science and Technology Application (Grant No. 2020-YF09-00048-SN), the Sichuan Science and Technology Program (Grant No. 2021JDRC0003), and the Innovation Fund of Postgraduate, Xihua University (Grant ycjj2020116).}

\markboth
{Chaojin Qing \headeretal: ELM-based Frame Synchronization in Nonlinear Distortion Scenario Using Superimposed Training}
{Chaojin Qing \headeretal: ELM-based Frame Synchronization in Nonlinear Distortion Scenario Using Superimposed Training}

\corresp{Corresponding author: Chaojin Qing (e-mail: qingchj@uestc.edu.cn).}

\begin{abstract}
The requirement of high spectrum efficiency puts forward higher requirements on frame synchronization (FS) in wireless communication systems. Meanwhile, a large number of nonlinear devices or blocks will inevitably cause nonlinear distortion. To avoid the occupation of bandwidth resources and overcome the difficulty of nonlinear distortion, an extreme learning machine (ELM)-based network is introduced into the superimposed training-based FS with nonlinear distortion. Firstly, a preprocessing procedure is utilized to reap the features of synchronization metric (SM). Then, based on the rough features of SM, an ELM network is constructed to estimate the offset of frame boundary. The analysis and experiment results show that, compared with existing methods, the proposed method can improve the error probability of FS and bit error rate (BER) of symbol detection (SD). In addition, this improvement has its robustness against the impacts of parameter variations.
\end{abstract}

\begin{keywords}
frame synchronization, extreme learning machine, nonlinear distortion, superimposed training
\end{keywords}

\titlepgskip=-15pt
\maketitle

\section{Introduction}
\IEEEPARstart{D}{ue} to the limited bandwidth resources, wireless communication systems pursue high spectrum efficiency in the past few decades \cite{Ref_1}. As we can see, the spectrum efficiency of the fifth generation (5G) wireless communication system is many times higher than that of the fourth generation (4G) wireless communication system \cite{Ref_2}, \cite{Ref_3}. In wireless communication systems, frame synchronization (FS) is a fundamental and essential task to guarantee the overall system performance \cite{Ref_4}, which usually consumes substantial bandwidth resources to overcome the synchronization challenge \cite{Ref_5}. Thus, during the FS phase, the contradiction between the high bandwidth resources consumption and the high spectrum efficiency requirement  needs to be resolved. Meanwhile, the wireless communication system has a large number of nonlinear devices or blocks, e.g., high power amplifier (HPA), digital to analog converter (DAC), etc., inevitably causing nonlinear distortion \cite{Ref_6}, \cite{Ref_7}. With limited considerations for nonlinear distortion, the classical methods (e.g., correlation-based FS \cite{Ref_8}) and the recent solutions (e.g., compressed sensing-based FS in \cite{Ref_9}) are usually difficult to apply in nonlinear distortion scenarios \cite{Ref_10}. Therefore, the FS is facing challenges from not only the spectrum efficiency but also the nonlinear distortion.

To cope with nonlinear distortion, machine learning (ML), in particular, deep learning (DL) has shown its prominent ability \cite{Ref_11}, \cite{Ref_12}. In recent years, DL has been applied in wireless communication, e.g., signal detection \cite{Ref_13}, precoding \cite{Ref_14}, channel state information (CSI) feedback \cite{Ref_15}, channel estimation \cite{Ref_16}, \cite{Ref_17}, mobile Internet of Things (IoT) \cite{Ref_38}, etc. However, these DL-based approaches exist weaknesses such as long-time training, complex parameter tuning \cite{Ref_10}, \cite{Ref_18} etc. Different from DL-based scheme, as a single hidden layer feed-forward neural network, extreme learning machine (ELM) can learn quickly, randomly generate for input
weight and hidden bias, require no gradient back-propagation, and has good generalization performance \cite{Ref_10}, \cite{Ref_19}, \cite{Ref_20}. As one of the effective options, using ELM to deal with nonlinear distortion is a promising solution.

For saving bandwidth resources and thus improving spectrum efficiency, the FS using superimposed training sequence is an attractive scheme. Without any occupation of bandwidth resources, this FS superimposes the training sequence on data symbols, yielding more transmission symbols than that of non-superposition mode \cite{Ref_21} in the same transmission interval. The superimposed training-based FS has been investigated in past years, e.g., \cite{Ref_21}--\cite{Ref_24}. These promising FSs promote us to develop further explorations, especially for the scenarios with nonlinear distortion.

Inspired by those advantages of ELM networks and superimposed training, we investigate an ELM-based FS by using superimposed training, which overcomes the challenges from spectrum efficiency and nonlinear distortion during the FS phase. In our work, the merits to cope with nonlinear distortion can be reaped by ELM networks, and then high spectrum efficiency can be achieved by using superimposed training. The combination of ELM network and superimposed training further improves the FS performance in the scenarios of nonlinear distortion, e.g., the error probability of FS. To our best knowledge, for ELM-based FS, there are limited works to focus on nonlinear distortion, much less to focus on superimposed training.

\subsection{RELATED WORKS}
We respectively present the related works of DL-based FS and ELM-based FS as follows.

The DL-based FS has been investigated in \cite{Ref_25}--\cite{Ref_28}. In \cite{Ref_25}, an artificial neural network (ANN)-based synchronization method was proposed. For the end-to-end communication systems, \cite{Ref_26} and \cite{Ref_27} investigated the FS based on neural network (NN). In \cite{Ref_26}, to achieve the task of FS, a deep neural network (DNN) was employed to auto-encoder, and a convolutional neural network (CNN) was developed in \cite{Ref_27} to compensate impairments introduced by timing offset and timing error of sampling. As for \cite{Ref_28}, a CNN-based FS method was proposed to convert the one dimension (1D) correlator to two dimension (2D) matrix to find the frame offset. From \cite{Ref_25}--\cite{Ref_28}, the DL technology provides effective approaches for FS. Nevertheless, these DL-based FSs are still challenged by many issues, such as long training time, complex parameter tuning, large memory requirements, etc.

Relative to DL networks, the ELM network features many advantages \cite{Ref_10}, \cite{Ref_19}, \cite{Ref_20}, e.g., the gradient back-propagation can be avoided, the output weight can be obtained by solving the least square (LS) question, and a single-hidden layer is employed for the feed-forward neural network, etc. In \cite{Ref_10}, an ELM-based time-division FS was proposed with the consideration of nonlinear distortion. In \cite{Ref_29}, the ELM network was employed to compensate the residual time offset with time-division mode. Although the error probability of FS is lower than conventional method, the training sequence for FS still occupies the bandwidth resources, which reduces systems' spectrum efficiency. To avoid the occupation of bandwidth resources, an ELM-based FS method using superimposed training is proposed in this paper to reduce the error probability of FS.

\subsection{CONTRIBUTIONS}

To overcome the challenges of spectrum efficiency and nonlinear distortion during the FS phase, the ELM-based FS using superimposed training is investigated in this paper. The main contributions of this paper are summarized as follows.

\begin{itemize}
  \item Firstly, an ELM-based FS method by using superimposed training is proposed. In contrast to the ELM-based time-division FS scheme in \cite{Ref_9}, not only the occupation of bandwidth resources is avoided in the proposed FS method, but also the smaller error probability of FS is achieved with the same energy cost.
  \item Secondly, the superimposed training-based FS is investigated in the scenarios of nonlinear distortion. Our investigation remedies the deficiencies of the existing superimposed training-based FS, which cannot work well in the scenarios of nonlinear distortion and is suitable for practical application.

  \item Thirdly, extensive experiments are given to verify the effectiveness of the proposed method in this paper. Compared with the classical correlation method in \cite{Ref_8} and the time division method in \cite{Ref_10}, both the FS's error probability and the symbol detection (SD)'s bit error rate (BER) are reduced with the same energy consumption. In addition, the proposed FS presents its robustness against the impacts of parameters.
\end{itemize}

The remainder of this paper is structured as follows: In Section II, we describe the system model. The ELM-based FS using superimposed training method is presented in Section III, followed by the experimental results and analysis are illustrated in Section IV. Finally, Section V concludes our work.

\textit{Notations}: Bold face upper case and lower case letters denote matrix and vector respectively. ${\left(\cdot \right)^T}$, ${\left(\cdot \right)^H}$, $\mathbf{(\cdot)^\dag}$, denote the transpose, conjugate transpose, and matrix pseudo-inverse, respectively. ${\left\|  \cdot  \right\|_2}$ is the Frobenius norm. $\left| {x} \right|$ denotes the absolute value of $x$ and $\left| {\mathbf{x}} \right|$ denotes the absolute value operation to the entry-wise of vector $\mathbf{x}$.

\section{SYSTEM MODEL}

At the receiver, the received $ M \times 1$ complex-valued signal vector, denoted as $\mathbf{y}$, can be expressed as
\begin{equation}
\label{eq:1}
{\mathbf{y}} = {\mathbf{H}}{\widetilde {\mathbf{x}}_{{\text{ext}}}} + {\mathbf{n}} ,
\end{equation}
where $\mathbf{n} \in \mathbb{C}^{M \times 1}$  represents the complex additive white Gaussian noise (AWGN) vector with zero-mean and $\sigma ^2$-variance entries. ${\widetilde {\mathbf{x}}_{{\text{ext}}}} \in {\mathbb{C}^{\left( {2N - L + 1} \right) \times 1}}$ denotes the extended vector of transmitted signal with nonlinear distortion, which can be given by

\begin{equation}
\label{eq:2}
{\widetilde {\mathbf{x}}_{{\text{ext}}}} = {\left[ {\underbrace {0, \cdots ,0}_\tau ,{{\widetilde {\mathbf{x}}}^T},\underbrace {0, \cdots ,0}_{N - L - \tau  + 1}} \right]^T} ,
\end{equation}
where $N$ and $L$ are the size of search window and the number of multi-path, respectively. The unknown frame boundary offset is denoted as $\tau $, whose range is $0 \le \tau  \leq N - {L} + 1$.
In (\ref{eq:2}), ${\mathbf{\widetilde x}} = {\left[ {{{\widetilde x}_1},{{\widetilde x}_2}, \cdots ,{{\widetilde x}_N}} \right]^T}$ denotes the distorted transmitted signal and can be expressed as

\begin{equation}
\label{eq:3}
{\mathbf{\widetilde x}} = {f_{\mathrm{dis}}}\left( {\mathbf{x}} \right) ,
\end{equation}
where ${f_{\mathrm{dis}}}(\cdot)$ represents the influence of nonlinear distortion. ${\mathbf{x}} = {\left[ {{x_1},{x_2}, \cdots ,{x_N}} \right]^T}$ is the superimposed transmitting signal without nonlinear distortion, which can be given by

\begin{equation}
\label{eq:4}
{\mathbf{x}} = \sqrt {\rho {E}} {\mathbf{s}} + \sqrt {\left( {1 - \rho } \right){E}} {\mathbf{c}} ,
\end{equation}
where $\rho  \in \left[ {0,1} \right]$ represents the power proportional coefficient (PPC), ${E}$ denotes the transmitted power of superimposed transmitting signal. ${\mathbf{s}} \in {\mathbb{C}^{N \times 1}}$ is the training sequence and ${\mathbf{c}} \in {\mathbb{C}^{N \times 1}}$ is the modulated data symbol.

The complex matrix ${\mathbf{H}} \in {\mathbb{C}^{M \times (2N - L + 1)}}$ given in (\ref{eq:1}) is an $M \times (2N-L+1)$ cyclic matrix, which can be defined as

\begin{equation}
\label{eq:5}
{{\mathbf{H}} = \left[ {\begin{array}{*{20}{c}}
  {{h_1}}&0& \cdots  \\
   \vdots &{{h_1}}& \ddots  \\
  {{h_L}}& \vdots & \ddots  \\
  0&{{h_L}}& \ddots  \\
   \vdots &0& \ddots  \\
  {}& \vdots & \ddots
\end{array}} \right]} ,
\end{equation}
where ${\mathbf{h}} = {\left[ {{h_1},{h_2},\cdots ,{h_L}} \right]^T}$ denotes the finite CIR vector of $L$ samples memory, and ${h_l}$ represents the complex-valued channel impulse response (CIR) of the $l$th path, $l = 1,2, \cdots ,L$.

With the received signal $\mathbf{y}$ given in (\ref{eq:1}), we employ ELM network to implement FS using superimposed training, as well as reducing the influence of nonlinear distortion.



\section{ELM-based FS}
The performance of FS is seriously degraded by the influence of nonlinear distortion. To conquer this difficulty, the ELM network is introduced into FS using superimposed training due to its prominent ability to cope with nonlinear distortion \cite{Ref_10}. In the following subsections, we first present the pre-processing of FS in Section III-\textit{A}. Then, the ELM-based FS method is given in Section III-\textit{B}.

\subsection{PRE-PROCESSING FOR FS}
In wireless systems, the FS is usually difficult to obtain, especially for nonlinear distortion scenarios. In particular, the error probability of DL-based timing synchronization is far higher than that of matched filtering in \cite{Ref_25}. Similar behaviors are also observed in ELM-based FS experiments. Thus, a pre-processing of FS is employed to coarsely capture the features of SM. According to \cite{Ref_8}, by using the cross-correlation based method, the SM vector ${\mathbf{g}} \in {\mathbb{R}^{N \times 1}}$ can be expressed by
\begin{equation}
\label{eq:6}
{\mathbf{g}}{\text{ = }}{\left| {{{\mathbf{S}}^H}{\mathbf{y}}} \right|^2} .
\end{equation}
Here, the $M \times N$ complex matrix ${\mathbf{S}}$ can be written as
\begin{equation}
\label{eq:7}
{\mathbf{S}} = \left[ {\begin{array}{*{20}{c}}
  {{s_1}}&0& \cdots  \\
   \vdots &{{s_1}}& \ddots  \\
  {{s_N}}& \vdots & \ddots  \\
  0&{{s_N}}& \ddots  \\
   \vdots &0& \ddots  \\
  {}& \vdots & \ddots
\end{array}} \right],
\end{equation}
where ${s_i},i = 1,2, \cdots ,N$, represents the $i$th entry of training sequence ${\mathbf{s}}$. Need to be mentioned that, besides the cross-correlation based SM in (\ref{eq:6}), other SMs can also be applied in our method with the similar processing.

In order to standardize the training of ELM network, the ${\mathbf{g}}$ given in (\ref{eq:6}) is normalized as

\begin{equation}
\label{eq:8}
\overline {\mathbf{g}}  = {{{\mathbf{g}} \mathord{\left/
 {\vphantom {{\mathbf{g}} {\left\| {\mathbf{g}} \right\|}}} \right.
 \kern-\nulldelimiterspace} {\left\| {\mathbf{g}} \right\|}}_2} .
\end{equation}

With the normalized SM (i.e., $\overline {\mathbf{g}}$), an ELM network is utilized to conquer nonlinear distortion and improve SMs for superimposed training-based FS, which is described in the following subsection.

\subsection{ELM-Based FS}
The ELM-based network is employed to improve SMs and decrease the influence of nonlinear distortion. The ELM-based FS includes offline and online procedures, which are elaborated in TABLE \ref{table_I} and TABLE \ref{table_II}, respectively. The offline training procedure is described as follows.

\subsubsection{OFFLINE TRAINING SPECIFICATION}
The offline procedure is elaborated in TABLE~\ref{table_I}. In the following, we first describe the data collection for ELM-net training.
\
\begin{itemize}
\item{}
\textit{Data Collection For ELM-Net training}
\end{itemize}

For ELM-net training, ${N_t}$ samples of input signals and offset labels, denoted by $\left\{ {\left( {{{\overline {\mathbf{g}} }_i},{{\mathbf{T}}_i}} \right)} \right\},i = 1,2, \cdots ,{N_t}$ are collected to form a  training set. According to the ${N_t}$ collected ${\overline {\mathbf{g}} _i}$ forms the input matrix $\overline {\mathbf{g}} \in {\mathbb{R}^{N \times {N_t}}}$, which can be written as

\begin{equation}
\label{eq:9}
\overline {\mathbf{g}}  = \left[ {{{\overline {\mathbf{g}} }_1},{{\overline {\mathbf{g}} }_2}, \cdots ,{{\overline {\mathbf{g}} }_{{N_t}}}} \right] .
\end{equation}

Similarly, ${N_t}$ offset label vectors ${\mathbf{T}} _i$ are converted to construct the target output matrix $\mathbf{T} \in \mathbb{R}^{N \times N_t}$, which can be given by

\begin{equation}
\label{eq:10}
{\mathbf{T} = \left[ \mathbf{T}_1, \mathbf{T}_2, \cdots,\mathbf{T}_{N_t}\right]} ,
\end{equation}
where the label ${{\mathbf{T}}_i}$ can be encoded according to one-hot mode, i.e.,

\begin{equation}
\label{eq:11}
{{\bf{T}}_i} = {\left[ {\underbrace {0, \cdots ,0}_{{\tau _i}},1,\underbrace {0, \cdots ,0}_{N - {\tau _i} - 1}} \right]^T} ,
\end{equation}
where ${\tau _i}$ is the $i$th sample's frame boundary offset.

In this paper, the input weights $\mathbf{W} \in {\mathbb{R}^{\widetilde{N} \times {N}}}$ and hidden layer biases $\mathbf{b} \in \mathbb{R}^{\widetilde{N} \times {1}}$ of ELM network (with $\widetilde N$ hidden neuron number) are respectively randomly chosen, which is similar to the standard process of ELM network \cite{Ref_20}. It should be noted that the initial values of $\mathbf{W}$ and $\mathbf{b}$ have a certain impact on the FS performance, we mainly investigate the FS method in this paper. Admittedly, initial values can further improve the FS's error probability and SD's BER. Then, the input weights $\mathbf{W}$ and hidden layer biases $\mathbf{b}$ are saved in storage space for later use in the offline and online procedure.

\begin{itemize}
\item{}
\textit{Networking Training}
\end{itemize}

As shown in TABLE~\ref{table_I}, during the training procedure, the training data-set $\left\{ {\left( {\overline{\mathbf{g}},\mathbf{T} } \right)} \right\}$, input wights $\mathbf{W}$ and hidden layer biases $\mathbf{b}$ are gradually loaded from storage space. Then, with $\left\{ {\left( {\overline{\mathbf{g}},\mathbf{T} } \right)} \right\}$, $\mathbf{W}$ and $\mathbf{b}$, the output matrix of the hidden layer ${\mathbf{H}} \in {\mathbb{R}^{\widetilde N \times {N_t}}}$ can be given by

\begin{equation}
\label{eq:12}
{\mathbf{H}} = \sigma \left( {{\mathbf{W}}\overline {\mathbf{g}}  + {\mathbf{b}}} \right) ,
\end{equation}
where $\sigma \left(  \cdot  \right)$ represents the activation function, e.g., sigmoid \cite{Ref_30}, hyperbolic tangent \cite{Ref_31}, rectified linear units (ReLU) \cite{Ref_32}, etc. The sigmoid is used in this ELM network \cite{Ref_20}. The objective of ELM network is actually to find the suitable output weights $\mathbf{\Upsilon}  \in {\mathbb{R}^{N \times \widetilde N}}$ to approximate the target output matrix ${\mathbf{T}}$, which can be expressed as \cite{Ref_19}

\begin{equation}
\label{eq:13}
{\mathbf{T}} = \mathbf{\Upsilon}{\mathbf{H}} ,
\end{equation}
where $\mathbf{\Upsilon} = \left[ {{{\mathbf{\Upsilon}}_1}, \cdots ,{{\mathbf{\Upsilon}}_k}, \cdots ,{{\mathbf{\Upsilon}}_{\widetilde N}}} \right]$, ${\mathbf{\Upsilon}}_k$ denotes the output weighting vector connecting the $k$th hidden neuron and the output neurons. The LS solution $\widehat {\mathbf{\Upsilon}} $ of ${\mathbf{T}} = \mathbf{\Upsilon}{\mathbf{H}}$ with minimum norm of output weights $\mathbf{\Upsilon}$ is given by \cite{Ref_19}

\begin{equation}
\label{eq:14}
\widehat {\mathbf{\Upsilon}}  = \mathop {\min }\limits_{\mathbf{\Upsilon}}  \left\| {\mathbf{\Upsilon} {\mathbf{H}} - {\mathbf{T}}} \right\| = {\mathbf{T}}{{\mathbf{H}}^\dag } ,
\end{equation}
the output weighting matrix $\mathbf{\Upsilon}$ is learned from offline training of ELM network, which is saved in storage space for online running.

\begin{table}[!ht]
\renewcommand\arraystretch{1.2}
\caption{TRAINING PROCEDURE}
\label{table_I}
\centering

\begin{tabular}{p{0.95\linewidth}}
\hline
\hline
\\

~~A training set $\left\{ {\left. {\left( {{{{\mathbf{\bar g }}}_i},{{\mathbf{T}}_i}} \right)} \right|i = 1,2, \cdots ,{N_t}} \right\}$ generated by SMs and one-hot label and these two are corresponding to the input and desired output of the ELM, respectively.
  \begin{enumerate}[step 1 :]
    \item Randomly chose the input weights $\mathbf{W} \in \mathbb{R}^{\widetilde{N} \times N}$ and hidden layer biases $\mathbf{b} \in \mathbb{R}^{\widetilde{N} \times 1}$, respectively, set the hidden neuron number $\widetilde{N}$.

    \item With ${\overline {\mathbf{g}}}$,  $\mathbf{W}$ and $ \mathbf{b}$, the output matrix of hidden layer  $\mathbf{H} \in \mathbb{R}^{\widetilde{N} \times N_t}$ is calculated through activation function $\sigma \left(  \cdot  \right)$ according to (\ref{eq:12}).

    \item Construct the target output matrix $\mathbf{T}$ using  offset label vector $\mathbf{T}_i$ according to (\ref{eq:10}), compute the output weights $\mathbf{\Upsilon} \in{\mathbb{R}^{{N} \times{\widetilde{N}}}}$ according to (\ref{eq:14}).
  \end{enumerate}\\
\hline
\hline

\end{tabular}
\end{table}

\subsubsection{ONLINE DEPLOYMENT}
With the learned ELM-based network parameters, the online running procedure can be implemented in this subsection, which is shown in TABLE~\ref{table_II}.

\begin{table}[!ht]
\renewcommand\arraystretch{1.2}
\caption{ONLINE PROCEDURE}
\label{table_II}
\centering

\begin{tabular}{p{0.95\linewidth}}
\hline
\hline
\\

  \begin{enumerate}[step 1 :]
    \item From (\ref{eq:6}) to (\ref{eq:8}), perform the preprocessing to obtain metric vector $\overline {\mathbf{q}}$ as the input of the trained ELM-net.

    \item With input metric vector $\overline {\mathbf{q}}$, the learned output weights $\mathbf{\Upsilon}$, the randomly chosen input weights $\mathbf{W}$ and hidden layer biases $\mathbf{b}$, the ELM network output $\mathbf{O}$ can be obtain according to (\ref{eq:15}).

    \item Find the location of the maximum square of the absolute value of the element in $\mathbf{O}$, (i.e., estimation of frame boundary offset $\widehat{\tau}$) according to (\ref{eq:16}).
  \end{enumerate}\\
\hline
\hline

\end{tabular}
\end{table}

With the input metric vector $\overline {\mathbf{q}} $, which can obtained by employing the preprocessing procedure according to (\ref{eq:6})-(\ref{eq:8}), the learned output weights $\mathbf{\Upsilon}$, the random input weights $\mathbf{W}$ and hidden layer biases $\mathbf{b}$, the ELM network output ${{\mathbf{O}}} \in {\mathbb{R}^{N \times 1}}$ can be written as

\begin{equation}
\label{eq:15}
{\mathbf{O}} = \mathbf{\Upsilon}  \cdot \sigma \left( {{\mathbf{W}}{{\overline {\mathbf{q}} }} + {\mathbf{b}}} \right) ,
\end{equation}
where ${\mathbf{O}} = {\left[ {{o_1},{o_2}, \cdots ,{o_N}} \right]^T}$, and the estimation of frame boundary offset can be expressed as

\begin{equation}
\label{eq:16}
\begin{aligned}
{\widehat{\tau}} = &\mathop {\arg \max }\limits_{1 \le j \le {N} } {|{o_{j}}|^2} .
\end{aligned}
\end{equation}

With the estimation from (\ref{eq:16}), the FS is completed by acquiring the frame's starting point $\widehat{\tau}$, After FS, the SD is performed, and the detected symbol $\widehat{\mathbf{c}}$ can be represented as

\begin{equation}
\label{eq:17}
{\widehat {\mathbf{c}} = \frac{{{\widetilde {\mathbf{x}}_{{\text{est}}}} - \sqrt {\rho {E}} {\mathbf{s}}}}{{\sqrt {1 - \rho } }}} ,
\end{equation}
where ${\widetilde {\mathbf{x}}_{{\text{est}}}}$ denotes the estimation of  superimposed transmitting signal in the scenario of nonlinear distortion, and can be obtained by

\begin{equation}
\label{eq:18}
{\widetilde {\mathbf{x}}_{{\text{est}}}} = {{\mathbf{H}}^\dag }{\mathbf{y}} .
\end{equation}

The FS and SD can be achieved according to (\ref{eq:6})-(\ref{eq:18}), which improve SMs and address the issues about multi-path interfere and nonlinear distortion.

\section{EXPERIMENTAL ANALYSIS}
In this section, numerical results of the proposed ELM-based FS using superimposed training are given. Firstly, basic parameters and definitions involved in simulations are given in Section IV-\textit{A}. Then, in Section IV-\textit{B}, the FS's error probability and SD's BER of the proposed scheme with nonlinear distortion is shown to verify the effectiveness of the proposed ELM-based FS, followed by the robustness of improvement with different parameters is discussed in Section IV-\textit{C}. At last, the computational time complexity is analyzed in Section IV-\textit{D}.

%

\Figure[t!](topskip=0pt, botskip=0pt, midskip=0pt)[width=3.3 in]{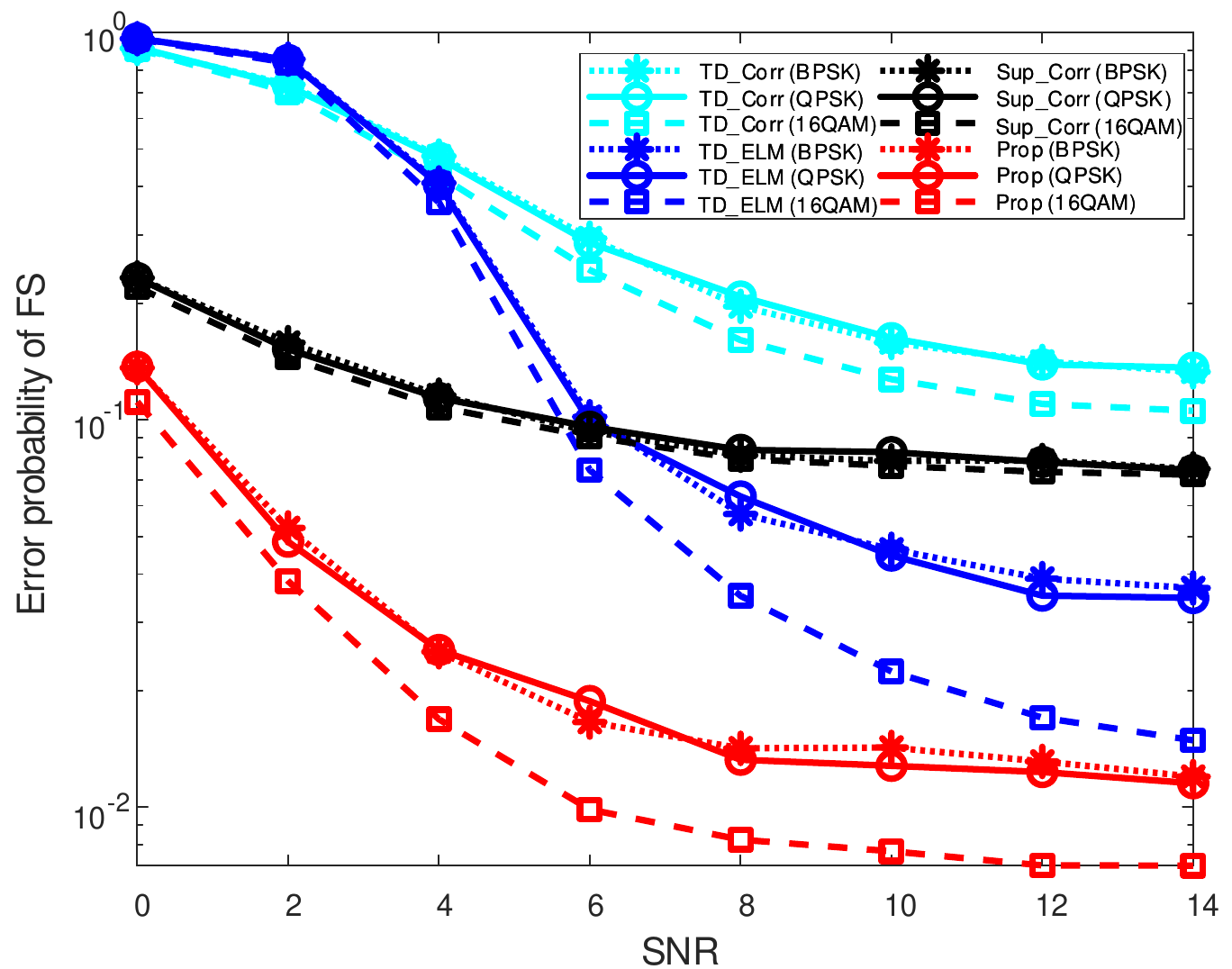}
{Error probability of FS vs. SNR, where $N = 512$, $L=8$ and $\operatorname{EVM}  = 35\% $.\label{fig1}}

\Figure[t!](topskip=0pt, botskip=0pt, midskip=0pt)[width=3.3 in]{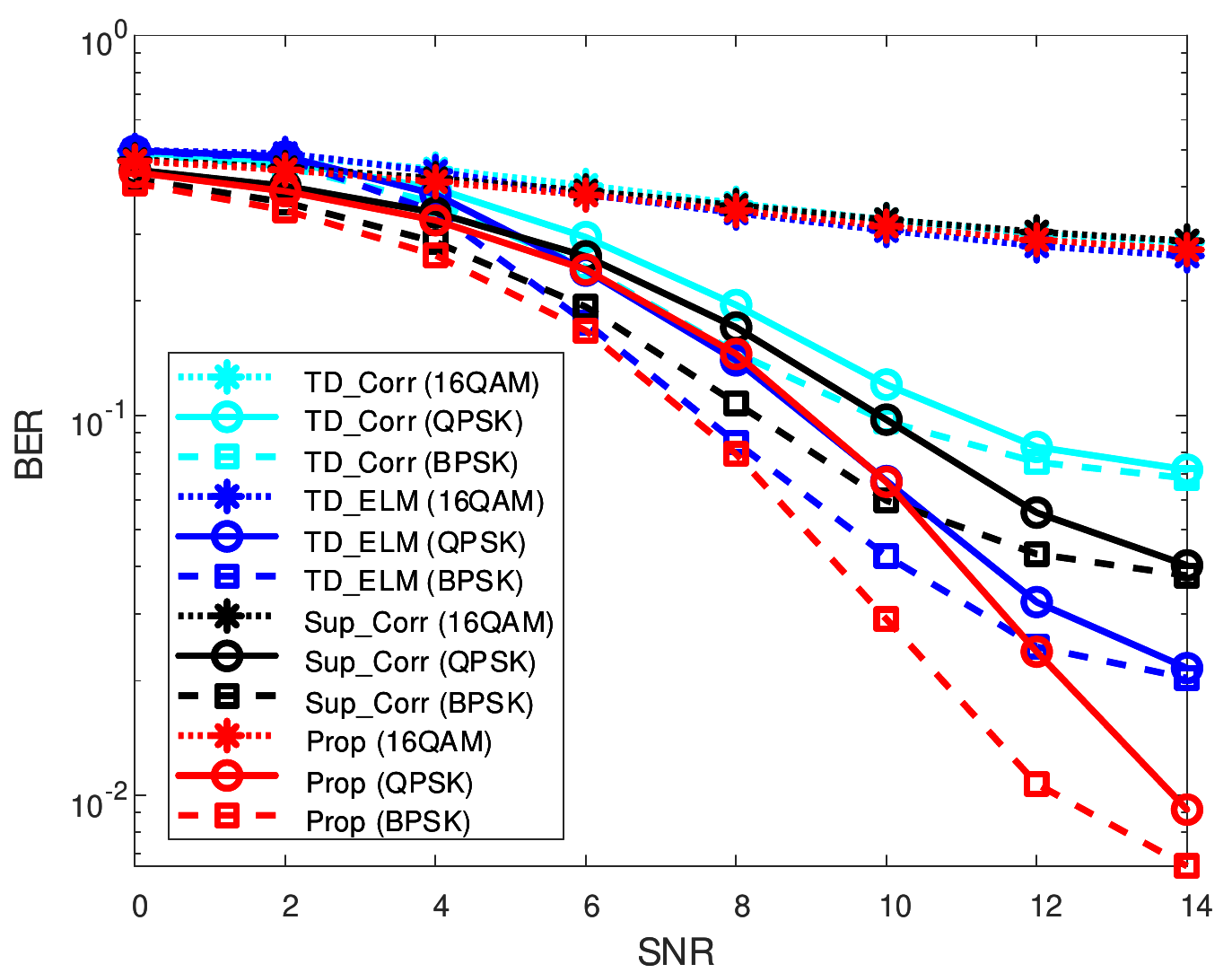}
{BER vs. SNR, where $N = 512$, $L = 8$ and $\operatorname{EVM}  = 35\% $.\label{fig2}}

\subsection{PARAMETER SETTING}

In the simulations, the basic parameters are set as $N = 512$, $M = 2N = 1024$, $\widetilde{N} = 10N = 5120$ \cite{Ref_19}, \cite{Ref_37}, $L = 8$, and ${N_t} = {10^5}$. The Zadoff-Chu sequence \cite{Ref_33} is employed as the training sequence $\mathbf{s}$. For the time-division method in \cite{Ref_10}, ${N_s} = 16$ is considered as the length of training sequence. By referencing \cite{Ref_21} and considering the total performance of FS's error probability and SD's BER, $\rho = 0.3$ is adopted in this paper. The modulated data symbol $\mathbf{c}$ is formed according to the symbol of quadrature-phase-shift-keying (QPSK) modulation. For the channel model, the multi-path Rayleigh fading channel with an exponentially-decayed power coefficient $\eta=0.2$ is considered, where each of the following $L - 1$ paths is set as zero-valued with a probability of 0.5 beside the first path to keep the same situation as \cite{Ref_9} and \cite{Ref_10}. For the sake of fair comparison, we assume the superimposed FS and the time-division FS consume the same energy for transmitting symbols.

\Figure[t!](topskip=0pt, botskip=0pt, midskip=0pt)[width=7 in]{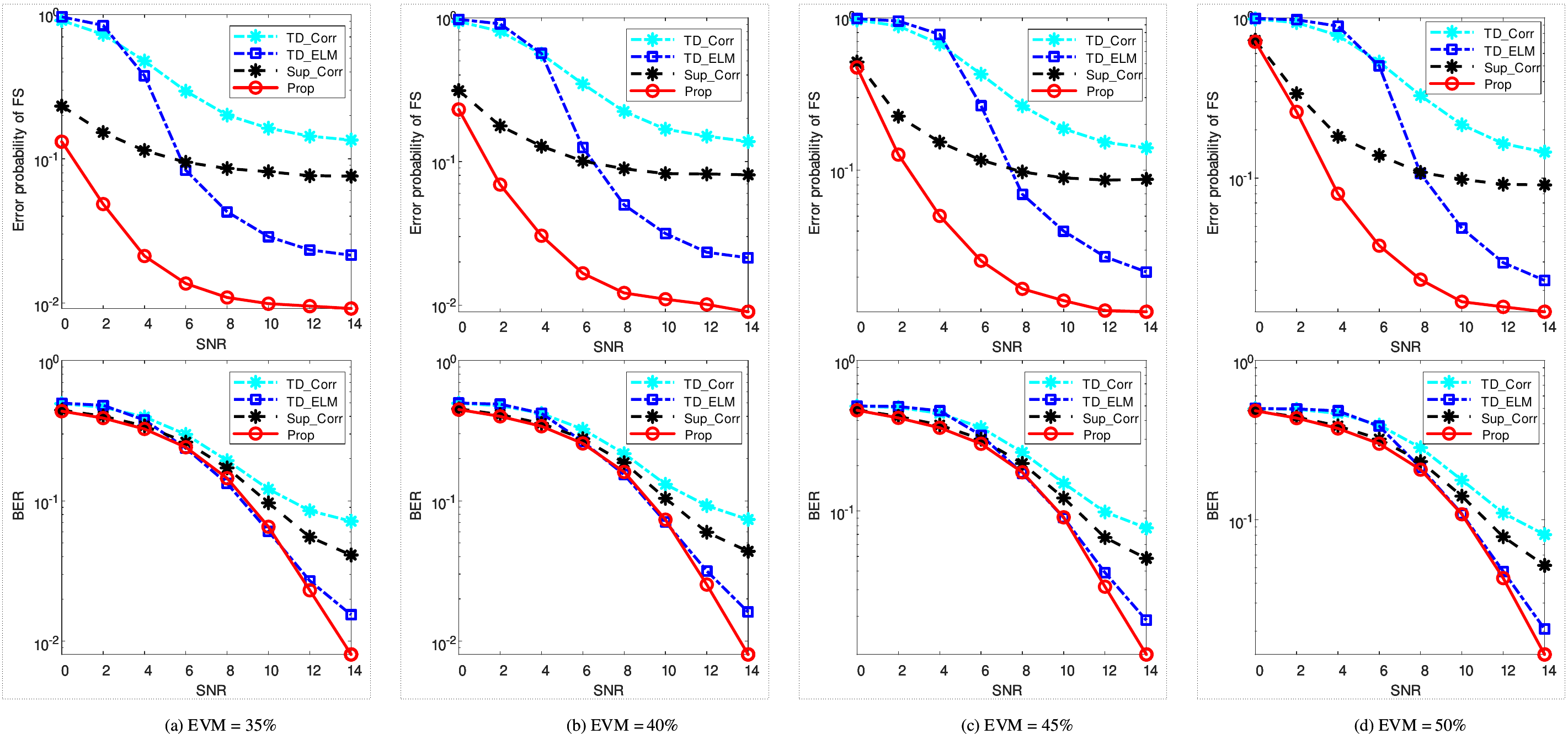}
{Error probability of FS and BER against the impact of EVM, where $\operatorname{EVM}  = 35\%$, $\operatorname{EVM}  = 40\%$, $\operatorname{EVM}  = 45\%$, and $\operatorname{EVM}  = 50\%$ are considered, respectively.\label{fig3}}

Definitions involved are listed as follows. The signal-to-noise ratio (SNR) in decibel (dB) is defined as \cite{Ref_34}
\begin{equation}
\label{eq:19}
SNR = 10{\log _{10}}\left( {\frac{{{E}}}{{{\sigma ^2}}}} \right) .
\end{equation}
For nonlinear distortion, the HPA effect is taken into account in these simulations. The nonlinear amplitude $A\left( x \right)$ and phase $\Phi \left( x \right)$ are obtained by \cite{Ref_35}

\begin{equation}
\label{eq:21}
A\left( x \right) = \frac{{{\alpha _a}x}}{{1 + {\beta _a}{x^2}}},{\rm{ }}\Phi \left( x \right) = \frac{{{\alpha _\phi }{x^2}}}{{1 + {\beta _\phi }{x^2}}},
\end{equation}
where ${\alpha _a} = 1.96$, ${\beta _a} = 0.99$, ${\alpha _\phi } = 2.53$, and ${\beta _\phi } = 2.82$ are considered in the experiments according to \cite{Ref_35}. To measure the distortion intensity, the error vector magnitude (EVM) is used in this paper, which is expressed as \cite{Ref_36}

\begin{equation}
\label{eq:21}
{\operatorname{EVM} \left( \%  \right) = \sqrt {\frac{{\sum\limits_{n = 1}^N {{{\left| {{{\widetilde x}_n} - {R_n}} \right|}^2}} }}{{\sum\limits_{n = 1}^N {{{\left| {{R_n}} \right|}^2}} }}}} ,
\end{equation}
where ${{\widetilde x}_n}$ is the $n$-th distorted symbol through HPA, which denotes the HPA workings in saturated region. ${R_n}$ denotes the desired linear outputs of HPA given the same input without amplification distortion. In this paper, the EVM is set as $\operatorname{EVM}  = 35\% $ except for the robustness analysis against EVMs.

For simplicity, ``Prop'', ``TD\_Corr'', ``TD\_ELM'', and ``Sup\_Corr'' are used to denote the proposed ELM-based superimposed FS, the correlation-based time-division FS in \cite{Ref_8}, the ELM-based time-division FS in \cite{Ref_10}, and the correlation-based superimposed FS method in \cite{Ref_21}, respectively.

\subsection{FS and SD PERFORMANCE}

To validate the effectiveness of the proposed ELM-based FS using superimposed training, the error probability of FS and BER of SD under different SNRs are illustrated in Fig. \ref{fig1} and Fig. \ref{fig2}, respectively.

The effectiveness of the error probability of FS is presented in Fig. \ref{fig1}. It could be observed that the ``Prop'' reaches the smallest error probability among different methods. This reflects the ``Prop'' obtains the best performance of error probability for FS, and thus can work well in the scenarios of nonlinear distortion. In addition, some insights of FS with nonlinear distortion can be achieved in Fig. \ref{fig1}. Firstly, the error probability of ``Sup\_Corr'' is smaller than that of ``TD\_Corr'' with the same energy consumption, which embodies the superiority of superimposed FS compared with time division FS. Secondly, for relatively high SNR (e.g., $SNR > 6$dB), the ``TD\_ELM'' has a smaller error probability of FS than those of ``Sup\_Corr'' and ``TD\_Corr''. That is, the ELM network effectively suppresses the nonlinear distortion for time division FS in \cite{Ref_8}, and the ELM-based time-division FS can further obtain better performance of error probability than that of superimposed FS in \cite{Ref_21}. Although the ELM-based time-division FS has shown its effectiveness to suppress the nonlinear distortion, it can only be effective in a relatively high SNR region, and the merits of superimposed FS have not been developed. The ``Prop'' develops the merits of superimposed training and ELM network, and thus obtains the smallest error probability of FS in all given SNR region. Therefore, the combination of superimposed training and ELM network in ``Prop'' possesses its effectiveness to improve the error probability of FS in the scenarios of nonlinear distortion.

\Figure[t!](topskip=0pt, botskip=0pt, midskip=0pt)[width=7 in]{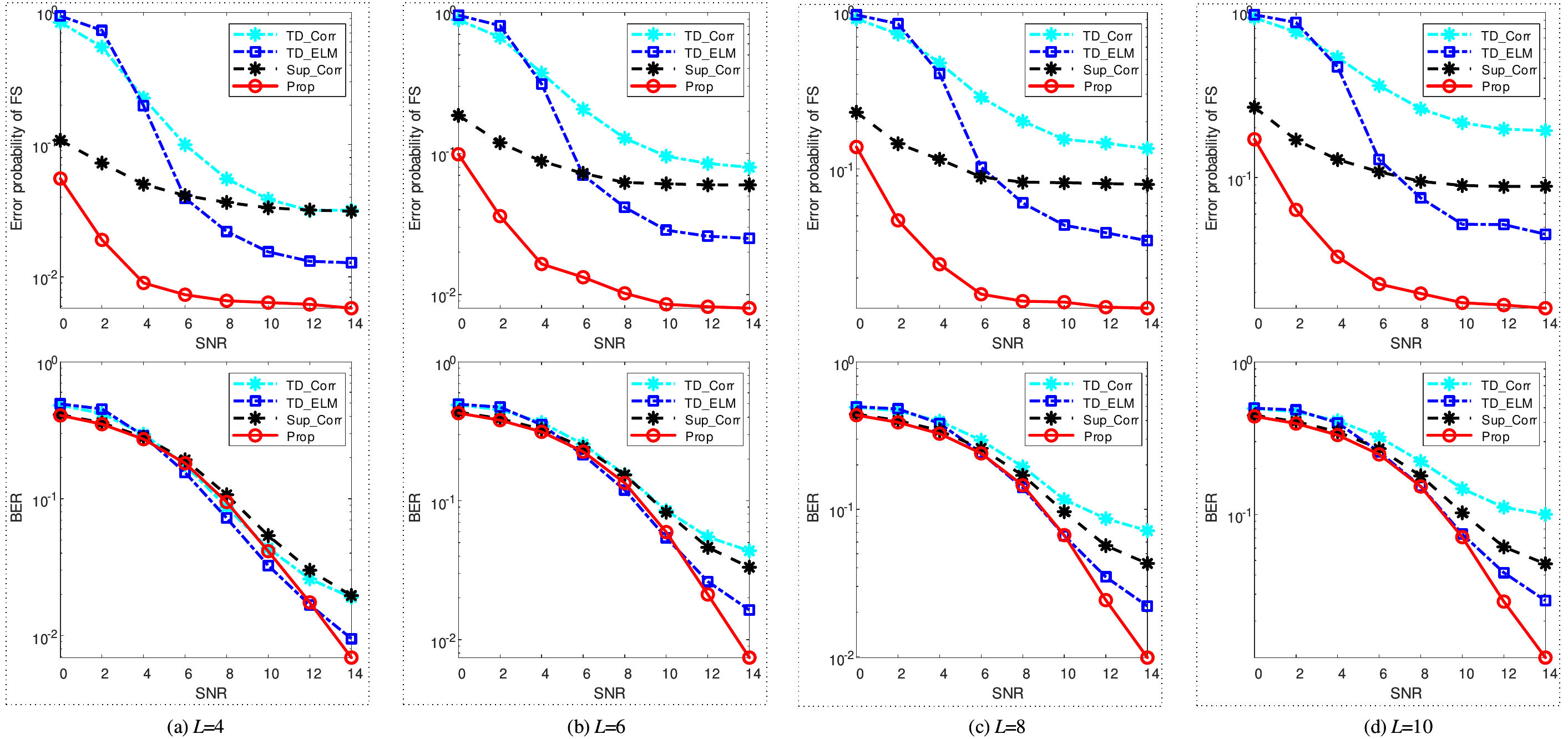}
{Error probability of FS and BER against the impact of $L$, where $L=4$, $L=6$, $L=8$, and $L=10$ are considered, respectively.\label{fig4}}

Since the training sequence $\mathbf{s}$ is superimposed on the modulated data symbol $\mathbf{c}$, it needs to be verified whether the superimposed interference (from the superimposed training) degrades the detection performance of data symbols. In this paper, the BER of SD is used to measure the detection performance and is plotted in Fig. \ref{fig2}. From Fig. \ref{fig2}, the ``Prop'' achieves the smallest BER for almost all given SNRs. Thus, for the same energy consumption, the superimposed interference does not degrade the BER performance. On the contrary, the BER of ``Prop'' effectively benefits from the proposed FS, especially for the relatively high SNR region (e.g., $SNR > 10$dB). That is, with simple processing of interference cancellation given in (\ref{eq:17}), the ``Prop'' achieves the best BER performance among all given FS methods. In particular, the advantages of superimposed training and ELM network can be separately demonstrated from Fig. \ref{fig2}. Without using superimposition approaches, ``TD\_ELM'' obtains smaller BER than that of ``TD\_Corr'', which reflects the effectiveness of ELM network to deal with nonlinear distortion. We can also observe that the BER of ``Sup\_Corr'' is smaller than that of ``TD\_Corr''. That is, superimposed training used in ``Sup\_Corr'' is useful to improve the BER performance of ``TD\_Corr''. Thus, by combining the superimposed training and ELM network, the BER performance is improved.

As a whole, compared with the ``TD\_Corr'', ``TD\_ELM'', and ``Sup\_Corr'', both the FS's error probability and the SD's BER in the scenarios of nonlinear distortion are improved by ``Prop''. Especially, compared with the ``TD\_Corr'' and ``TD\_ELM'', the ``Prop'' can obtain the chance to transmit more data symbols, and thus can further improve the spectrum efficiency. By the way, with different modulation conditions (e.g., BPSK and 16QAM), the proposed method still improves the FS's error probability and SD's BER from Fig. \ref{fig1} and Fig. \ref{fig2}.

\subsection{ANALYSIS OF PARAMETER IMPACT}
In this subsection, the robustness of the proposed scheme against parameter variation is analysed. The impact of EVM is first discussed, followed by the number of multi-path (i.e., $L$), the transmitted frame-length $N$, and the PPC $\rho$. It is worth noting that, besides the change of the impact parameter (i.e, EVM, $L$, $N$, and $\rho$), other basic parameters remain the same as those given in Section IV-\textit{A} during the simulations.

\subsubsection{IMPACT OF EVM}
EVM is usually used to measure the distortion intensity. To analyze the robustness of the proposed method against different distortion intensities, Fig. \ref{fig3} respectively plots the curves of FS's error probability and SD's BER with different EVMs (i.e, $\mathrm{EVM} = 35 \%$, $\mathrm{EVM} = 40 \%$, $\mathrm{EVM} = 45 \%$, and $\mathrm{EVM} = 50 \%$).

From Fig. \ref{fig3}, compared with those of ``TD\_Corr'', ``TD\_ELM'', and ``Sup\_Corr'', the ``Prop'' method achieves the smallest error probability for each given EVM. That is, relative to the existing methods, the proposed FS scheme still improves the FS's error probability against varying EVMs. With the increase of EVM, the FS's error probabilities for all curves in Fig. \ref{fig3} (i.e., ``TD\_Corr'', ``TD\_ELM'', ``Sup\_Corr'', and ``Prop'') increase due to the rise of distortion intensity. However, the FS's error probability of ``Prop'' is smaller than those of ``TD\_Corr'', ``TD\_ELM'', and ``Sup\_Corr'', especially for the high SNR region. This reflects the proposed scheme can improve the FS's error probability against varying EVMs.

For the SD, Fig. \ref{fig3} shows the BER of ``Prop'' is smaller than those of ``TD\_Corr'' and ``Sup\_Corr'' in almost all given SNR regions. Especially, in a relatively high SNR region, e.g., $SNR \geq 12$dB, we can observe the BER of ``Prop'' is smaller than those of ``TD\_Corr'', ``Sup\_Corr'', and ``TD\_ELM''. Thus, for different EVMs, the ``Prop'' achieves similar or better BER performance.

As a result, against the impact of EVM, the ``Prop'' possesses its robustness for improving FS's error probability and SD's BER.

\Figure[t!](topskip=0pt, botskip=0pt, midskip=0pt)[width=7 in]{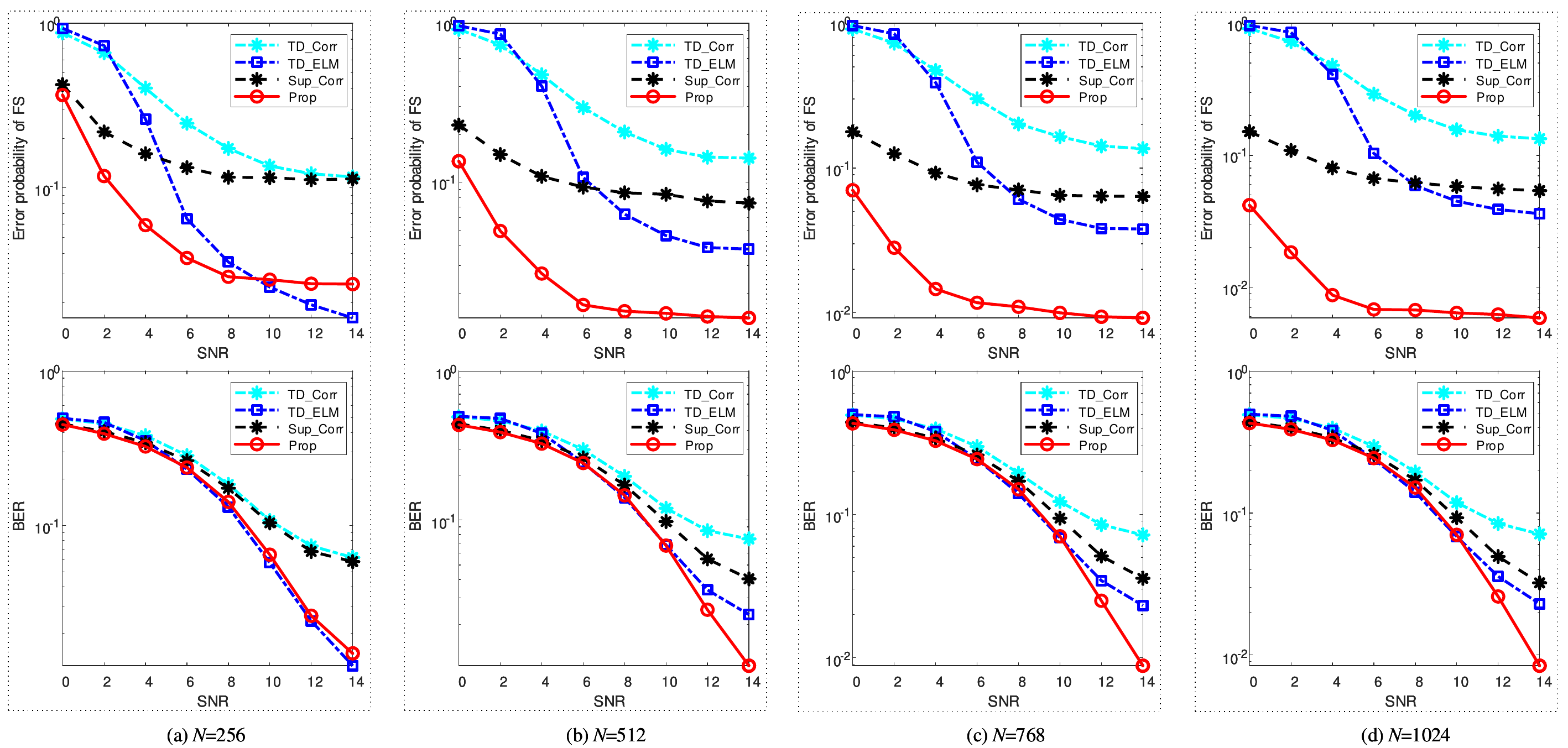}
{Error probability of FS and BER against the impact of $N$, where $N=256$, $N=512$, $N=768$, and $N=1024$ are considered, respectively.\label{fig5}}

\subsubsection{IMPACT OF $L$}

The FS's error probability and SD's BER are usually impacted by the number of multi-path (i.e., $L$). To demonstrate the robustness of the proposed FS scheme against the impact of $L$, the error probability of FS and BER of SD curves are given in Fig. \ref{fig4}, where $L = 4$, $L = 6$, $L = 8$, and $L = 10$ are considered, respectively.

From Fig. \ref{fig4}, relative to the ``TD\_Corr'', ``TD\_ELM'', and ``Sup\_Corr'', the ``Prop'' achieves the minimal error probability of FS for each given $L$. This reflects the ``Prop'' improves the FS's error probability of the existing methods with the variations of $L$. In addition, with the increase of $L$, the FS's error probabilities of ``Prop'', ``TD\_Corr'', ``TD\_ELM'', and ``Sup\_Corr'' rise with the enlargement of multi-path interference. Even so, the ``Prop'' still presents the ability to cope with the nonlinear distortion and multi-path interference under different values of $L$, and thus obtains the smallest FS's error probability. As a whole, against the impact of $L$, the ``Prop'' can robustly reduce FS's error probability.

From the curves of SD's BER in Fig. \ref{fig4}, the ``Prop'' achieves the smallest BER for relatively large $L$ (e.g., $L \geq 8$). This reflects the ``Prop'' possesses better BER performance compared with the existing methods when $L\geq8$. For the case where $4$dB$\leq SNR\leq12$dB and $L=4$, the BER of ``Prop'' is slightly higher than that of ``TD\_ELM''. This is because the ``TD\_ELM'' consumes additional bandwidth resources to avoid the superimposed interference of ``Prop''. Moreover, the ELM network is also employed by ``TD\_ELM'', and thus effectively suppresses the nonlinear distortion as well. Nevertheless, compared with  ``TD\_ELM'' for the case where $4$dB $ \leq SNR \leq $ $8$dB, the ``Prop'' has only slightly higher BER, while transmitting more data symbols to obtain higher spectrum efficiency. Besides, the BER of ``Prop'' is still smaller than that of ``TD\_ELM'' when $SNR\geq12$dB.

To sum up, against the impact of $L$, the ``Prop'' can effectively reduce the FS's error probability and SD's BER, especially for the cases of relatively large $L$ and relatively high SNR.

\subsubsection{IMPACT OF $N$}
Usually, the FS's error probability and SD's BER performance are influenced by the frame-length, i.e., $N$. To validate the robustness against the impact of $N$, the error probability of FS and BER of SD are illustrated in Fig. \ref{fig5} with different values of $N$ (i.e., $N = 256$, $N= 512$, $N= 768$, and $N = 1024$).

From Fig. \ref{fig5}, for the cases where $N=512$, $N=768$, and $N=1024$, the ``Prop'' obtains a smaller FS's error probability than those of ``TD\_Corr'', ``TD\_ELM'', and ``Sup\_Corr''. This reflects the ``Prop'' could reduce the error probability of FS against the varying $N$. When $N=256$ and $SNR\leq8$dB, the ``Prop'' can still obtain the minimum error probability of FS to embodies the robustness against the impact of $N$. However, for $SNR \geq10$dB, such situation could not be held. As can be seen in Fig. \ref{fig5}(a),  the FS's error probability of ``Prop'' is higher than that of ``TD\_ELM''. This is because the ``TD\_ELM'' can also suppress the nonlinear distortion as that of the  ``Prop'', while the relatively short $N$ brings ``Prop'' the difficulty to combat the superimposed interference from the data symbol $\mathbf{c}$. Even so, compared with the ``TD\_ELM'', the ``Prop'' saves the bandwidth resources and significantly reduces FS's error probability in a relatively low SNR region (e.g., $SNR \leq8$dB). In particular, relative to the ``TD\_ Corr'' and ``Sup\_Corr'', the ``Prop'' clearly reduces the error probability of FS for all given SNRs. In addition, with the increase of $N$, the FS's error probability of ``Prop'' decreases. This reflects the elongated $N$ can effectively suppress the superimposed interference of the ``Prop'', due to the elongation of the superimposed training sequence $\mathbf{s}$.

Compared with ``TD\_ELM'', ``TD\_Corr'', and ``Sup\_Corr'', the ``Prop'' has similar or smaller BER, which demonstrates the ``Prop'' obtains the similar or better SD's BER performance with different values of $N$.  Relative to ``TD\_Corr'' and ``Sup\_Corr'', the ``Prop'' reduces the BER for each given $N$, especially in relatively high SNR region (e.g., $SNR\geq10$dB). Meanwhile, for the case where $N=256$ and $0$dB$ \leq SNR \leq$ $10$dB, the ``Prop'' obtains similar BER performance as that of ``TD\_ELM''. When $SNR\geq10$dB, the BER of ``Prop'' is slightly higher than that of ``TD\_ELM''. The reasons are given as follows. On the one hand, the nonlinear distortion is also suppressed by employing the ELM network in ``TD\_ELM''.
On the other hand, the ``TD\_ELM'' consumes additional bandwidth to avoid superimposed interference. Moreover, the corresponding FS's error probability of ``TD\_ELM'' is smaller than that of ``Prop'', which also deteriorates the SD's BER performance. Even so, the ``Prop'' can save the bandwidth resource and obtains similar BER performance as that of ``TD\_ELM'' when $N = 256$. Especially, for relatively large $N$ (e.g., $N \geq 512$), the ``Prop'' can achieve lower BER than that of ``TD\_ELM'' when $SNR \geq 12$dB.

On a whole, with the varying of $N$, the ``Prop'' can improve the performance of FS's error probability and SD's BER, and this improvement possesses good robustness.

\begin{table}[]
\renewcommand{\arraystretch}{1}
\caption{Time complexity comparison}
\label{table_III} \centering
\begin{tabular}{|c|c|c|}
\hline
                    & TD\_ELM & Proposed \\ \hline
Training time       & about 19.6 minutes         & about 20.2 minutes     \\ \hline
Online running time & about 11.6 minutes         & about 11.7 minutes      \\ \hline
\end{tabular}
\end{table}

\subsection{COMPLEXITY ANALYSIS}

The training time and online running time between the correlation-based superimposed FS method in \cite{Ref_10} (i.e., ``TD\_ELM'') and the proposed ELM-based superimposed FS (i.e., ``Prop'') are illustrated in TABLE~\ref{table_III} to compare the computational time complexity.

For a fair comparison, $10^{5}$ experiments are conducted for ``Prop'' and ``TD\_ELM'' on the same server with Intel Xeon(R) E5-2620 CPU 2.1GHz\_16 by using Matlab software, respectively. During the experiments, only running time is considered, (i.e., the time to generate data sets and through channel is not included). From TABLE~\ref{table_III}, in the training phase, the ``TD\_ELM'' consumes about 19.6 minutes, while the ``Prop'' costs about 20.2 minutes. In the online running stage, the ``TD\_ELM'' consumes about 11.6 minutes, and the ``Prop'' costs about 11.7 minutes. We can see that the average training and online running time of ``Prop'' is slightly more than that of ``TD\_ELM'' in each experiment, while the ``Prop'' improves the spectrum efficiency with similar or better performance of FS's error probability and SD's BER relative to the existing methods.

\section{CONCLUSION}
In this work, we integrated superimposed training-based FS and ELM network to investigate an ELM-based FS scheme using superimposed training in nonlinear distortion scenarios. Firstly, a preprocessing procedure is employed to coarsely reap the features of SM. Then, an ELM network is introduced to conquer the impact of nonlinear distortion and to obtain the estimation of frame boundary offset. Compared with some existing methods, the proposed method can improve the error probability of FS and BER of SD, and those improvements are robust against parameter variation. In future works, we will investigate the generalization method of ELM-based FS to alleviate the difference between the data set from simulation and the data set in real scenarios.


\begin{IEEEbiography}[{\includegraphics[width=1in,height=1.25in,clip,keepaspectratio]{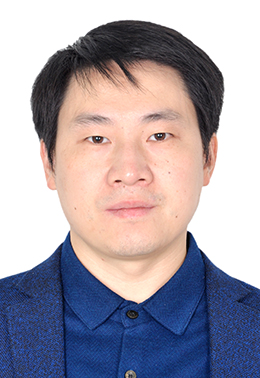}}]{Chaojin Qing} (M'15) received the B.S. degree in communication engineering from Chengdu University of Information Technology, Chengdu, China, in 2001, the M.S. and Ph.D. degrees in communications and information systems from the University of Electronic Science and Technology of China, Chengdu, China, in 2006 and 2011, respectively. From November 2015 to December 2016, he was a Visiting Scholar with Broadband Communication Research Group (BBCR) of the University of Waterloo, Waterloo, ON, Canada.

From 2001 to 2004, he was a teacher with the Communications Engineering Teaching and Research Office, Chengdu University of Information Technology, Chengdu, China. Since 2011, he has been a Professor with the School of Electrical Engineering and Electronic Information, Xihua University, Chengdu, China. He is the author of more than 50 papers and more than 20 Chinese inventions. His research interests include detection and estimation, massive MIMO systems, and deep learning in physical layer of wireless communications.
\end{IEEEbiography}

\begin{IEEEbiography}[{\includegraphics[width=1in,height=1.25in,clip,keepaspectratio]{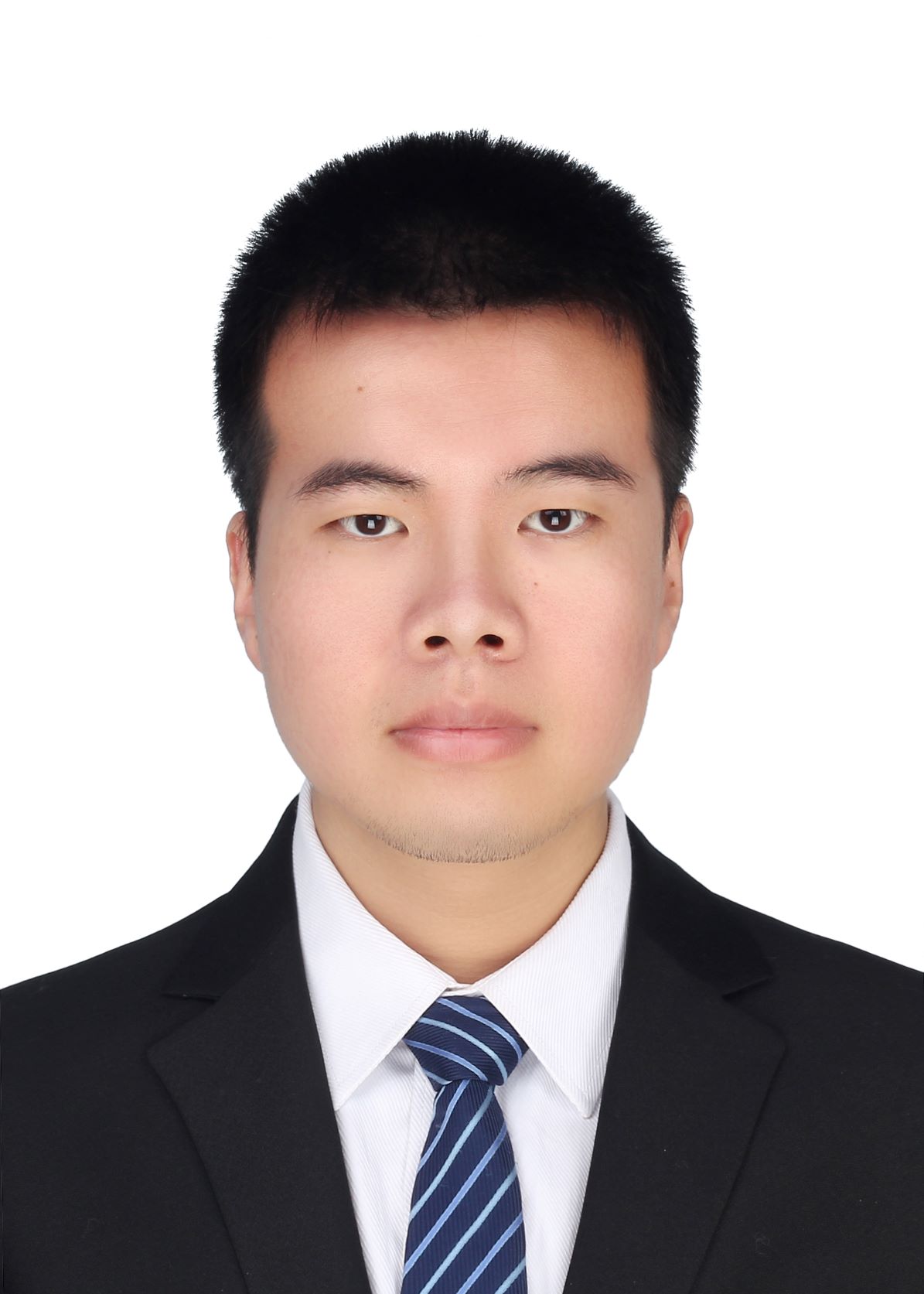}}]{Wang Yu}
received the B. S. degree from the School of Electrical Engineering and Electronic Information, Xihua University, Chengdu, China, in 2017, where he is currently  pursuing the M. S. degree under the supervision of Prof. Qing. His research interests include frame synchronization and frequency synchronization in wireless communications, and deep learning in physical layer of wireless communications.
\end{IEEEbiography}

\begin{IEEEbiography}[{\includegraphics[width=1in,height=1.25in,clip,keepaspectratio]{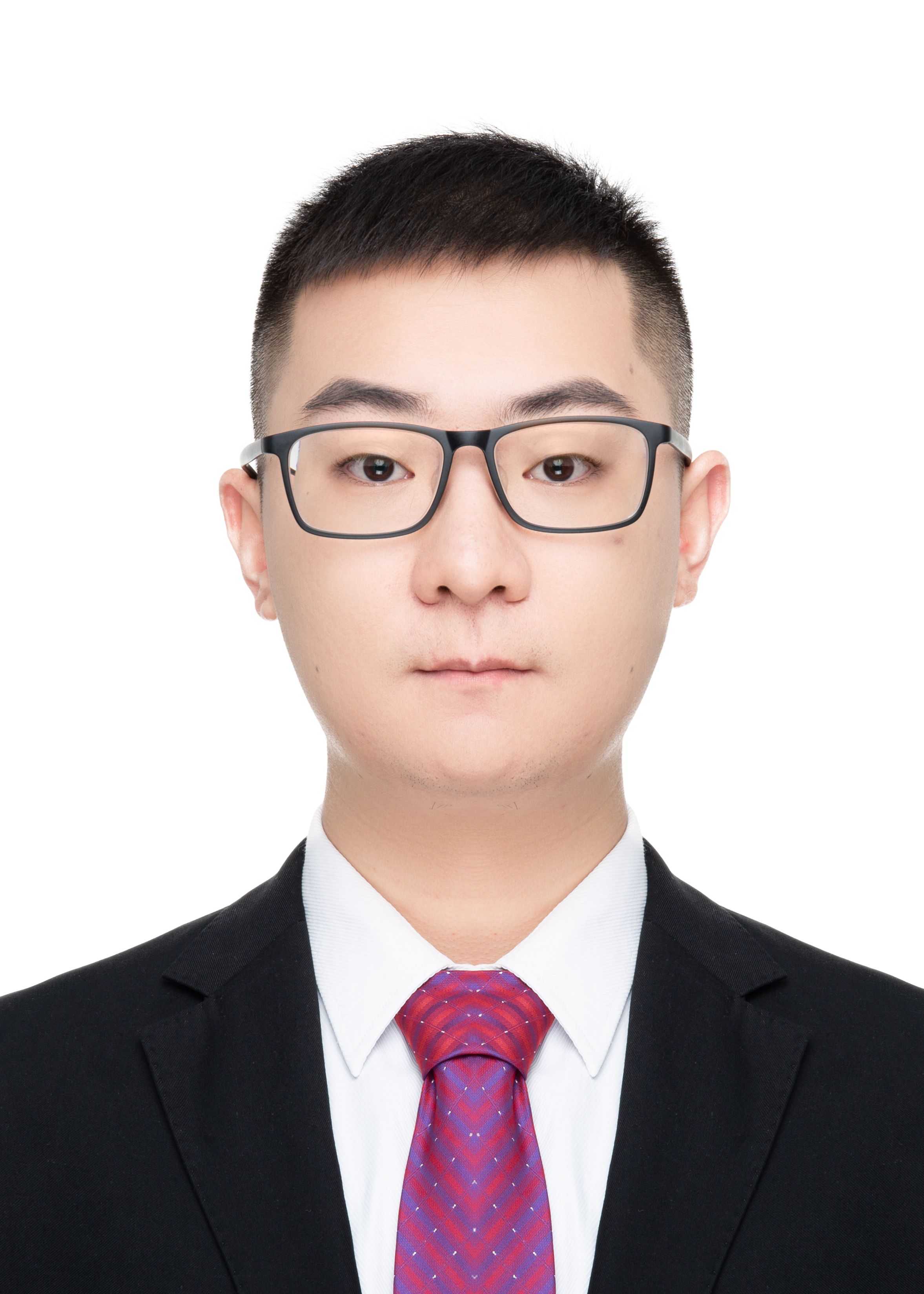}}]{Shuhai Tang}
received the B. S. degree from the School of Electrical Engineering and Electronic Information, Xihua University, Chengdu, China, in 2020, where he is currently pursuing the M. S. degree under the supervision of Prof. Qing. His research interests include synchronization and channel estimation in OFDM system, and machine learning applications in physical layer of wireless communications.
\end{IEEEbiography}

\begin{IEEEbiography}[{\includegraphics[width=1in,height=1.25in,clip,keepaspectratio]{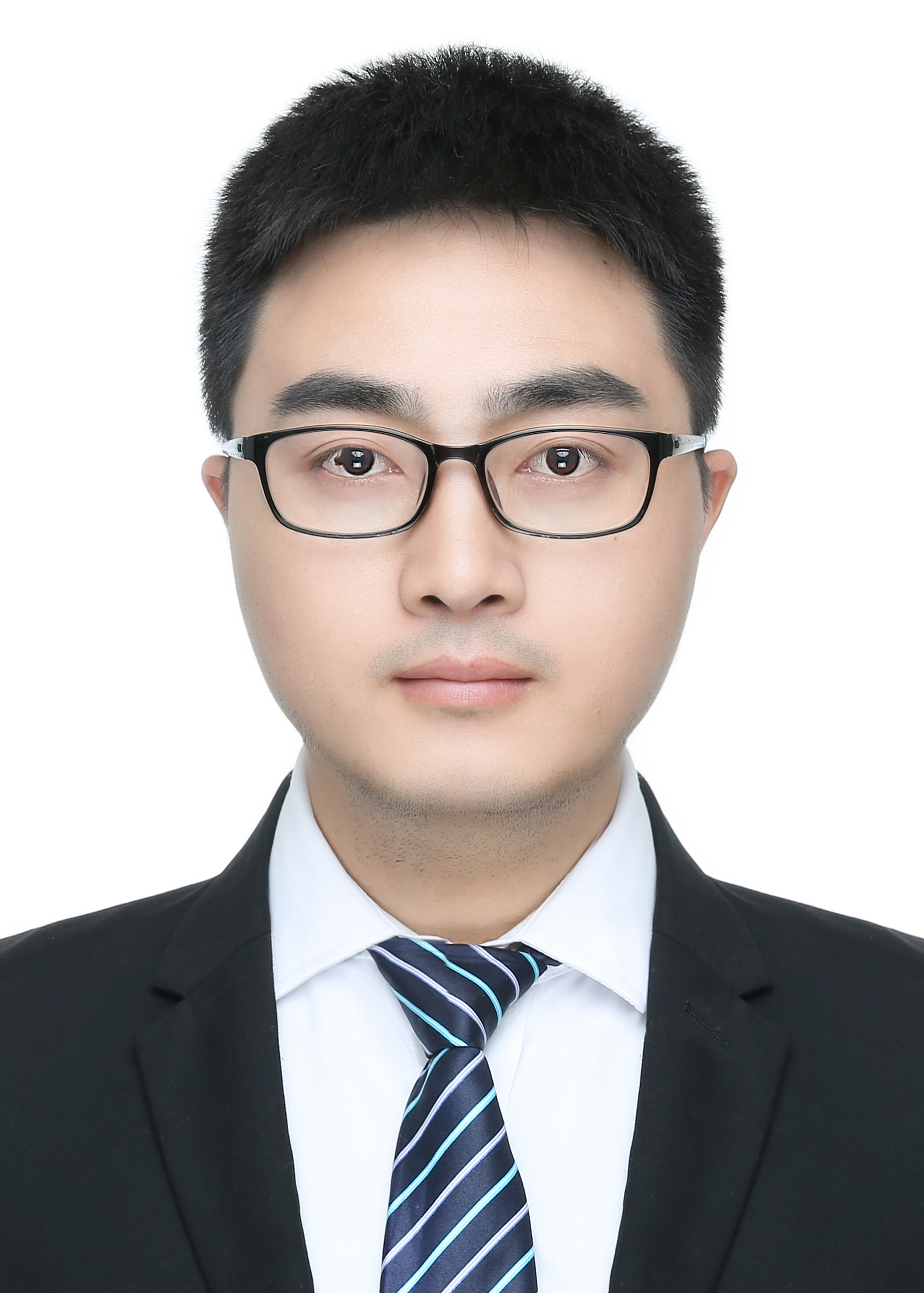}}]{Chuangui Rao}
received B. S. degree in Electrical Engineering and Information from Southwest Petroleum University, China, in 2017. He is currently pursuing the M. S. degree with the School of Electrical Engineering and Electronic Information, Xihua University, Chengdu, China, under the supervision of Prof. Qing. His research interests include frame synchronization and channel estimation, and deep learning applications in physical layer of wireless communications.
\end{IEEEbiography}

\begin{IEEEbiography}[{\includegraphics[width=1in,height=1.25in,clip,keepaspectratio]{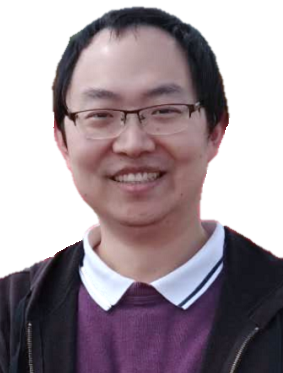}}]{Jiafan Wang} (S'15) received his B.S. degree and M.S. degree in Electrical Engineering from University of Electronic Science and Technology of China in 2006 and 2009, respectively. He accomplished the Ph.D.  degree in Computer Engineering at Texas A$\&$M University, College Station, TX, USA in 2017.

His major field of study is smart integrated circuit design, which includes multi-dimensional non-deterministic gate implementation with systematic optimization framework, self-training Analog/Digital Mixed-System for circuit feature calibration, and configurable locking mechanism against Analog IP piracy. He is currently working in Synopsys Inc. to develop the world's leading silicon chip design software in electronic design automation (EDA) industry.
\end{IEEEbiography}

\EOD

\end{document}